\documentclass{ifacconf}

\newcommand{\problemtwo}{
  \renewcommand{\theproblem}{2($i, \varepsilon$)}
}
\usepackage{amsmath,amsfonts}
\usepackage{amssymb}
\usepackage{latexsym}
\usepackage{bm}
\usepackage{mathtools}
\usepackage{empheq}
\usepackage{xcolor}
\usepackage{algorithm}

\usepackage{graphicx}  
\usepackage[numbers]{natbib}
\usepackage{algorithmic}
\usepackage{array}
\usepackage{longtable}
\usepackage{float}
\usepackage{wrapfig}
\usepackage{enumerate}
\usepackage{multirow}
\usepackage{comment}
\usepackage{listings}
\usepackage{empheq}
\usepackage{color}
\usepackage{subcaption}
\usepackage{caption}
\usepackage{tikz}
\usepackage{optidef}
\usepackage{url}

\newtheorem{problem}{Problem}

\begin{document}
\begin{frontmatter}

\title{Gig-work Management System with Chance-Constraints Verification Algorithm} 

\author[First]{Kazuyoshi Fukuda} 
\author[First]{Masaki Inoue} 
\author[First]{Riko Asanaka}

\address[First]{Keio University(e-mail: kazu150207@keio.jp).}

\begin{abstract}
This paper proposes the framework of an efficient gig-work management system. A gig-work management system recommends one-off tasks with information about task hours and wages to gig-workers. To enable effective management, this paper develops a model of gig-workers' decision-making. Then, based on the model, we formulate an optimization problem to determine the optimal task hours and wages. The formulated problem belongs to the class of chance-constrained model predictive control (CC-MPC) problems. To efficiently solve the CC-MPC problem, we develop an approximate solution algorithm with guaranteed confidence levels.
Finally, we develop gig-worker models based on data collected through crowdsourcing.
\end{abstract}

\begin{keyword}
Gig-Work Management, Verification Algorithm, Nonlinear System Identification, Chance-Constrained Optimization, Discrete Event Modeling and Simulation
\end{keyword}

\end{frontmatter}

\section{Introduction}\label{intro}
In recent years, the advancement of digital platforms has led to the global expansion of a new employment style known as gig-work. Gig-work refers to a new work style in which workers undertake short-term or one-off tasks without forming long-term employment relationships with specific companies. It has become widespread in sectors such as food delivery and ride-sharing \cite{Timee,DeStefano2016, Kassi2018}. While the gig-work style offers flexibility to workers, it also poses challenges for platforms and employers in determining appropriate task hours and  wages \cite{HallKrueger2018}. 

The objective of this paper is to propose a systematic design of a gig-work management system (Gig-WMS) that optimizes task hours and wages \cite{AsanakaInoue}. To design Gig-WMS,  modeling the decision-making process of workers is essential. Several previous studies have addressed modeling of general workers, beyond the scope of gig-workers, as in the literature \cite{RouieYossi,PasupaSuzuki,WileyDeckroJackson,YumbeKomoda,LiuZhan,LinzSemykina,CammanLawler}. The study \cite{RouieYossi} states that the prediction of future wages plays an essential role in determining workers' decision-making on whether to accept a task. The study \cite{PasupaSuzuki} proposes a mathematical model describing the relationship between worker productivity and wages. Most of the previous studies have assumed deterministic decision-making models, where workers follow instructions with certainty. In contrast, gig-workers' decision-making is probabilistic. This paper presents a probability model for representing workers' decision-making by using the logit model, widely studied in behavioral economics \cite{McFadden}. Based on the logit model, this paper formulates a chance-constrained model predictive control (CC-MPC), which is studied in the literature \cite{Cannon2009,Blackmore2010,WangBoyd2010}, to dynamically determine task hours and wages. CC-MPC is a control method that predicts future scenarios based on probability models and iteratively computes optimal control actions while satisfying constraints. Many studies address the method of handling chance-constraints in CC-MPC, including robust reformulations \cite{ZhangCheng}, distribution-based soft constraint approaches \cite{OldewurtelMorari}, and scenario-based methods \cite{Schildbach2015}. In this paper, we develop an approximate solution algorithm with guaranteed confidence levels for CC-MPC based on the study \cite{AlamoTempo}. 

The remainder of this paper is structured as follows. 
Section~\ref{overview} provides an overview of Gig-WMS.
Section~\ref{model} presents the probabilistic decision-making model of gig-workers. 
Section~\ref{cont} formulates the CC-MPC problem of optimizing the task hours and wages based on the decision-making model. Furthermore, Section~\ref{cont} presents the solution method with satisfying chance constraints with guaranteed confidence levels.
Section~\ref{simu} presents numerical experiments that identify the gig-workers' decision-making model and verify the effectiveness of Gig-WMS. 
Finally, Section~\ref{conclu} gives concluding remarks.

\section{Overview of Gig-WMS}\label{overview}
\begin{figure}[t]
\centering
    \includegraphics[scale=0.3]{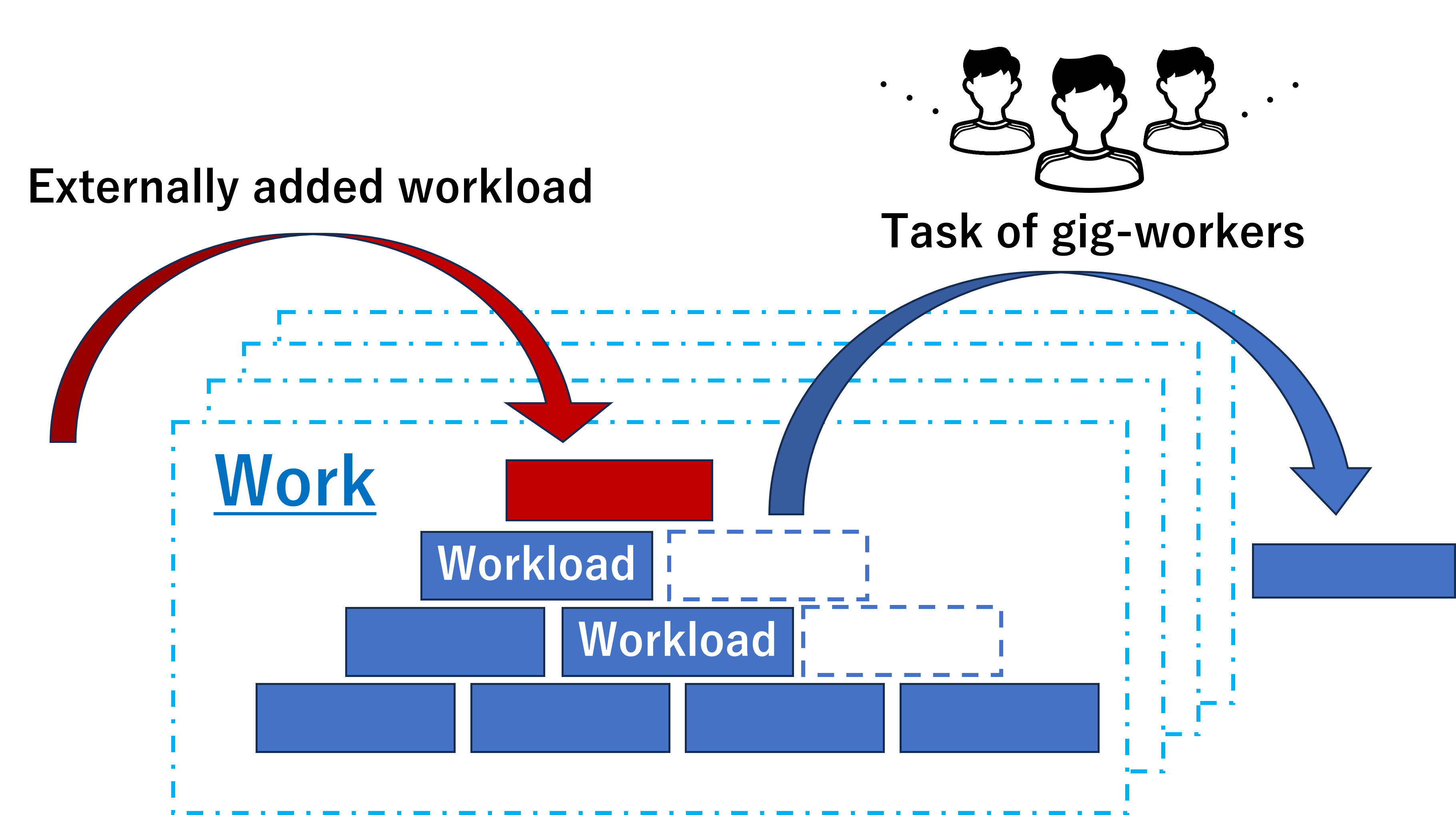}
    \caption{The transition of workload by task execution} \label{work_dyna}\vspace{-1mm}
\end{figure}

This section provides an overview of Gig-WMS. 
Within Gig-WMS, $m$ types of work are available, and $n$ gig-workers are participating. Each piece of work still carries a substantial workload to be reduced through the execution of individual tasks. 
The Gig-WMS platform functions as an intermediary, linking tasks to gig-workers. Task requests issued by employers include information on expected task hours and wages.  Based on the requests, gig-workers decide whether to accept the task. The task proceeds collectively through their efforts, and the remaining workload is reduced over time. Since workloads are continuously added from external sources, the cycle of short-term contracting and task execution is repeated continuously. This iterative process, summarized in Fig. \ref{work_dyna}, continues until the remaining workload falls below a predefined target threshold.

\begin{figure}[t]
\centering
    \includegraphics[scale=0.6]{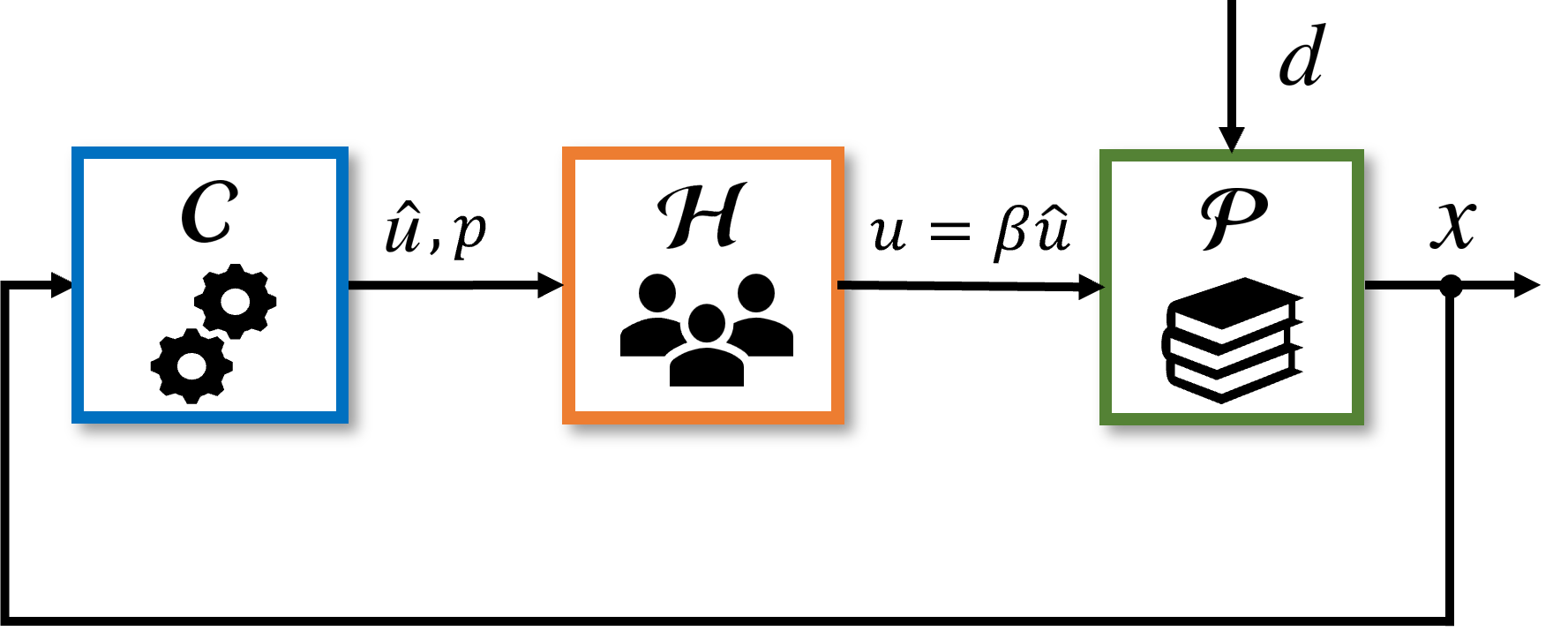}
    \caption{The block diagram of Gig-WMS} \label{flow}\vspace{-1mm}
\end{figure}

The block diagram of Gig-WMS is illustrated in Fig. \ref{flow}.
In Gig-WMS, $x$, $\hat{u}$, $p$, $u$, and $d\in \mathbb{R}^{m}$ represent workload, requested task hours, wages, actual task hours, and externally sourced workload, respectively. Symbol $\beta\in\{0,1\}$ represents acceptance of the requested task by some of the $n$ registered gig-workers. If at least one gig worker accepts the requested task, $\beta = 1$. If no one accepts it, $\beta = 0$.
The symbols used in this paper are summarized in Table \ref{table1}.
\begin{table}[t]  
 \caption{Definition of Symbols}  
 \label{table1}  
 \centering  
  \begin{tabular}{cl}  
   \hline  
   Symbol & Definition \\  
      \hline \hline  
    $n$ & Number of gig-workers \\ 
    $m$ & Number of works \\
    $x$ & Workload \\  
    $\hat{u}$ & Requested task hours \\  
    $u$ & Actual task hours \\  
    $p$ & Wage \\  
    $d$ & Externally added workload \\  
    $\beta$ & Acceptance of task by the gig-worker group \\  
    \hline  
  \end{tabular}  
\end{table}

In Fig. \ref{flow}, $\mathcal{H}$ represents the group of gig-workers,  $\mathcal{P}$ represents the dynamics of the workload, and $\mathcal{C}$ represents the controller.  
At each time step $k$, the controller $\mathcal{C}$ determines the task hours $\hat{u}(k)$ and the wages $p(k)$ based on the current remaining workload $x(k)$. The controller sends a task offer $(\hat{u}(k), p(k))$ to the gig-worker group. Each gig-worker decides whether to accept the task based on the task offer $(\hat{u}(k), p(k))$. If one or more gig-workers accept the task, that is, if $\beta(k)$ = 1, the remaining workload $x(k)$ is updated to $x(k+1)$ according to the tasks and the externally added workload $d(k)$. Based on updated $x(k+1)$, the controller $\mathcal{C}$ determines the next task hours $\hat{u}(k+1)$ and wages $p(k+1)$. By repeating the sequence, Gig-WMS aims to manage the workload so that it does not become excessively large. 

Typical examples of gig-work services include short-term delivery tasks or agricultural support tasks in such cases, the time step $k$ roughly corresponds to a daily unit, with task hours $\hat{u}$ typically ranging from 3 to 7 hours and wages $p$ about 1,351 JPY per hour in Japan \cite{Uber}. 

\section{Modeling}\label{model}
In this section, we derive the mathematical models for the dynamics of the workload $\mathcal{P}$ and the task acceptance probability in the decision-making of the gig-worker group $\mathcal{H}$. 
Subsection~\ref{Hmodel} presents the modeling of the gig-worker group's decision-making $\mathcal{H}$. Subsection~\ref{Pmodel} presents the modeling of workload dynamics $\mathcal{P}$.  

\subsection{Model of gig-workers $\mathcal{H}$}\label{Hmodel}

This subsection introduces the mathematical model of decision making $\mathcal{H}$, which is refined from the previous study by the authors \cite{AsanakaInoue}.

The task acceptance of each gig-worker $i$ is represented by a binary random variable $\beta_i \in \{0,1\}$, where $\beta_i = 1$ indicates task acceptance and $\beta_i = 0$ indicates the rejection. The probability of acceptance is modeled using a discrete choice model based on the worker's utility function. Specifically, the task acceptance probability is given by 
\begin{equation}\label{dec}  
\Pr({\beta}_i = 1) = \frac{1}{1 + \exp(-V_i)},  
\end{equation}  
where $V_i$ is the utility function of gig-worker $i$ and is defined by
\begin{equation} \label{utility}
V_i( \hat{u}, p,\nu_i) = \kappa\hat{u} + \lambda p + \nu_i.  
\end{equation}  
Based on the study \cite{TverskyKahneman}, we let $\kappa < 0$ and $\lambda > 0$. 
The term $\nu_i$ captures individual-specific factors that affect the worker's decision, and it varies across gig-workers. 

By substituting the utility function (\ref{utility}) into the acceptance probability model (\ref{dec}), we obtain 
\begin{equation}\label{dec_utility}
\Pr(\beta_i = 1) = \frac{1}{1 + \exp(-V_i( \hat{u}, p,\nu_i))}.  
\end{equation}  
Then, this implies that the probability of that at least one gig-worker in the group accepts the task is modeled by  
\begin{equation}\label{dec_all} 
\Pr(\beta = 1) = 1 - \prod_{i=1}^{n} \left( \frac{1}{1 + \exp\left( V_i( \hat{u}, p,\nu_i) \right)} \right).  
\end{equation}  
Once the decisions are made by the gig-workers, the actual task hours for the assigned gig-worker are described by
\begin{equation}\label{u}
u = \beta\hat{u}.
\end{equation}

This subsection was devoted to a decision-making model for gig-workers.
It is worth noting that behavioral differences across individuals or groups are characterized by the coefficients $\kappa$,$\lambda$, and $\nu_i$ in (\ref{utility}).  The parameters of the coefficients are estimated based on data collected through a crowd-sourced survey in Section~\ref{simu}.

\subsection{Model of works $\mathcal{P}$}\label{Pmodel} 

Recall that $\mathcal{P}$ represents a dynamics in which the remaining workload $x$ evolves based on the actual task hours $u$ by gig-workers and externally added workload $d$. The transition of the remaining workload $x$ is described by the following state equation  
\begin{equation} \label{dyna}
x(k + 1) =  A x(k) - u(k) + d(k),  
\end{equation}  
where $A \in \mathbb{R}^{m \times m}$ is a constant matrix representing how much the workload grows over time if no task is performed. As a case study, consider an inventory management. In cases where tasks are not completed by gig-workers, the inventory level increases continuously. This dynamic is captured by modeling the system $\mathcal{P}$ such that the system matrix $A$ has eigenvalues greater than or equal to 1.


It should be noted that, as seen in (\ref{dec_all}),(\ref{u}) and (\ref{dyna}), $\mathcal{P}$ represents a dynamical system subject to uncertainty in the control action $u(k)$. Such a class of plant systems are not typically seen in traditional optimal control or optimal management problems, including energy, vehicle, electrical devices, and so on \cite{Hjungle}. Even in robust control frameworks, it is uncommon for action terms to be treated probabilistically \cite{Tsai}.

\section{Controller design}\label{cont}
In this section, we first design CC-MPC based on the decision-making model, presented in Section~\ref{model}, in Subsection~\ref{CC-MPC}. Next, in Subsection~\ref{alg}, we develop an efficient approximate solution algorithm for the CC-MPC with the desired confidence level by applying the feasibility verification algorithm proposed in \cite{AlamoTempo}. 


\subsection{Formulation of CC-MPC}\label{CC-MPC}

This subsection aims to formulate the problem to optimize the requested task hours $\hat{u}$ and wages $p$ recommended to the entire gig-worker group for designing Gig-WMS. Since the decision-making model varies among gig-workers, we first optimize them for each gig-worker $i$ individually. Thus, we formulate a CC-MPC for each gig-worker $i$. Among the $n$ obtained optimal solutions, the one with the minimum wages is selected as the optimal solution.  To simplify notation, we let $\hat{U}_{N} = \{\hat{u}(k), \hat{u}(k+1),\ldots, \hat{u}(k +N - 1) \}$ and $p_{N} = \{p(k), p(k+1),\ldots, p(k +N - 1) \}$ to represent the time series of $\hat{u}$ and $p$, respectively.
 
\begin{problem}\label{prb1}  
Given $x(k)$, solve the following optimization problem to obtain ${\hat{U}}_{N}, P_{N}$  
\end{problem}  

\begin{minie}
  {\hat{U}_{N}, P_{N} \in \mathbb{R}_{+}^{N}} 
  {\sum_{t=k}^{k+N-1} p(t)}                  
  {\label{prb}}                              
  {}                                         
  \addConstraint{\Pr\bigl(x(k + N) > x_{\text{ref}}\bigr)\leq \eta} 
  \addConstraint{x(t + 1)= A x(t) - \beta(t)\hat{u}(t) + d(t)}
  \addConstraint{\Pr(\beta(t) = 0) \leq \varepsilon}
  \addConstraint{\forall t \in \{k,\dots,k+N-1\}}
\end{minie}

The objective function (7a) represents the total salary from the current time $k$ up to the prediction horizon $N$. The inequality (7b) represents a chance constraint that ensures the predicted remaining workload $x(k + N)$ stays below the desired threshold $x_\text{ref}$. The equation (7c) represents the plant dynamics. 
The inequality (7d) serves as a chance constraint on the task acceptance probability.

A sufficient condition for (7d) is given by 
\begin{equation} \label{ineq1}
\Pr(\beta_i(t) = 0) \leq \varepsilon^{\frac{1}{n}}. 
\end{equation}
This (\ref{ineq1}) is reduced to the  following inequality 
\begin{equation} \label{ineq2}
\begin{aligned}
\hat{u}(t) &\leq -\frac{1}{\kappa}\left[\lambda p(t) + \nu_i - \ln\left(\varepsilon^{-\frac{1}{n}} - 1\right) \right] \\
           &=: \bar{u}(p(t); \nu_i, \varepsilon)
\end{aligned}
\end{equation}
The following proposition holds for chance-constraint 
in (7d) and its corresponding deterministic constraint in
(\ref{ineq2}).


\begin{prop}\label{prp1}
Let $\hat{u}$ and $p$ satisfy (\ref{ineq2}). Then, (7d) holds for the same $\hat{u}$ and $p$.
\end{prop} 

According to \textit{Proposition~\ref{prp1}}, you can replace (7d) by (9) in \textit{Problem~\ref{prb1}} to obtain the feasible solution to the problem.

\subsection{Solution Algorithm}\label{alg}
Recall that \textit{Problem~\ref{prb1}} is an optimization problem that includes the chance-constraints (7b) and (7d).
Replacing (7d) with the deterministic constraint given in (9) does not lead to a straightforward solution of the resulting optimization problem, due to the presence of the other chance constraint (7b).
In this subsection, we address an efficient approximate solution algorithm for \textit{Problem~\ref{prb1}} by applying the feasibility verification algorithm proposed in \cite{AlamoTempo}.
The solution algorithm consists of the following three steps.\smallskip

{Step 1)
\textit{Problem~\ref{prb1}} is approximated by a deterministic formulation that remains dependent on $\varepsilon$, which governs the approximation accuracy.}

{Step 2) 
Obtain the solution to the approximation problem and verify whether it satisfies the original \textit{Problem~\ref{prb1}.}

{Step 3) 
Repeat 1) and 2) while adjusting $\varepsilon$ until the verification is completed.
}\smallskip

Each step is explained in detail below.


Step 1) Approximation of \textit{Problem~\ref{prb1}}: In this step, we derive a deterministic optimization problem as an approximation of \textit{Problem~\ref{prb1}}. In particular, we replace the chance constraint given in (7b) by a deterministic one. To this end, recall that $\varepsilon$ is sufficiently small in (7d), then it follows that $\Pr(\beta = 0)=0$. Therefore, it approximately follows that $\beta=1$ and \textit{Problem~\ref{prb1}} is reduced to the following deterministic problem.

\problemtwo
\begin{problem}\label{prb2}
Given $x(k)$, solve the following optimization problem to obtain ${\hat{U}}_{N}, P_{N}$
\end{problem}

\begin{minie}
    {{\hat{U}}_{N}, P_{N} \in \mathbb{R}_{+}^{N}}          
    { \sum_{t=k}^{k+N-1}  p(t)\label{eqprb-obj}} 
    {\label{prb}}  
    {} 
    \addConstraint{x(k + N) \leq x_{\text{ref}} } 
    \addConstraint{x(t + 1)= Ax(t) - \hat{u}(t) + d(t)} 
    \addConstraint{\hat{u}(t) \leq  \bar{u}(p(t) ; \nu_i(t), \varepsilon)}
    \vspace{3mm}  
    \addConstraint{\forall t \in \{k,\cdots,k+N-1\}}. 
\end{minie}
\textit{Problem~\ref{prb2}} is an approximation of \textit{Problem~\ref{prb1}}. 
\textit{Problem 2} is dependent on $i$ and $\varepsilon$. Therefore, we let \textit{Problem 2} be \textit{Problem~\ref{prb2}}.
The difference between the optimal solutions of the two problems diminishes when $\varepsilon$ is sufficiently small. We let the optimal solution of \textit{Problem~\ref{prb2}} be a candidate solution of \textit{Problem~\ref{prb1}} written as $(\hat{u}_c, p_c)$. 
 


Step 2) Verification of Feasibility: In this step, we verify that the candidate solution $(\hat{u}, p)$ is the feasible solution to \textit{Problem~\ref{prb1}} with confidence level $\delta \in (0, 1)$. The verification is performed by the method presented in the study \cite{AlamoTempo}. 

For the verification, 
we define a flag function $g(s,\hat{u}, p; \varepsilon)$ to evaluate the candidate solution $(\hat{u}, p)$ as 
\begin{equation}\label{binary}
g(s,\hat{u}, p; \varepsilon) =
\begin{cases}
     1 & \text{if $s$ holds,} \\
     0 & \text{otherwise.}
\end{cases}
\end{equation}

By repeating this evaluation and summing the outputs, we estimate the probability that $s$ holds.

Further, we let $m_l$ and $M_l$ be the level function and cardinal function of iteration number $l$, respectively defined as follows
\begin{equation}\label{m}
m_l = \lfloor bl \rfloor,
\end{equation}
\begin{equation}\label{verification1}
M_l = \left\lceil \frac{1}{\eta} \left( m_l + \ln \left( \frac{\zeta(\alpha) l^\alpha}{\delta} \right) + \sqrt{2m_l \ln \left( \frac{\zeta(\alpha) l^\alpha}{\delta} \right)} \right) \right\rceil,
\end{equation}
where $\zeta(\alpha)=\sum_{k=1}^{\infty} \frac{1}{k^\alpha}$ is the Riemann zeta function with $\zeta(2) = \pi^2/6$.
Then, using $m_l$ and $M_l$ and applying \textit{Corollary 2} in \cite{AlamoTempo} to our problem setup, we obtain the following proposition.

\begin{prop}\label{prp2}
Given confidence level $\delta \in (0, 1)$.  Suppose that
\begin{equation}\label{verification2}
\sum_{j=1}^{M_l}[1-g^{j}((10b),\hat{u}, p; \varepsilon)] \leq m_l 
\end{equation}
holds.  
Then, it holds that 
\begin{equation}\label{verification3}
\Pr\left\{g({{(7b)},\hat{u}, p; \varepsilon)=1}\right\} > 1-\delta.
\end{equation}
\end{prop}

Here, $g^j(s,\hat{u}, p; \varepsilon)$ represents the result of the $j$-th trial of the binary function $g(s,\hat{u}, p; \varepsilon)$, 
which determines whether the constraint $s$ is satisfied when $\hat{u}$ and $p$ are given as inputs.
Based on \textit{Proposition~\ref{prp2}}, we design the solution algorithm for \textit{Problem~\ref{prb1}}. If the candidate solution fails to satisfy (\ref{verification2}), it cannot be guaranteed to be feasible. In that case, we update $\varepsilon_{l}$ and solve \textit{Problem~\ref{prb2}} for $\varepsilon_{l+1}$. This requires feedback to reduce the gap between \textit{Problem\ref{prb1}} and \textit{Problem~\ref{prb2}}.

Step 3) Adjust $\varepsilon$: To increase the probability of passing the verification, $\varepsilon$ should be reduced. Thus, we define the update rule
\begin{equation}\label{update}
\varepsilon_{l+1}=\gamma \varepsilon_{l}.
\end{equation}
Based on (\ref{update}), we update $\varepsilon$  and solve  \textit{Problem~\ref{prb2}} again. We then re-verify the chance-constraint for the candidate solution. By repeating 2) and 3), the approximate feasible solution can be obtained.

\subsection{Summary of the Solution Algorithm}
The solution algorithm for \textit{Problem~\ref{prb1}} is summarized in \textbf{Algorithm~\ref{alg1}}.

\begin{algorithm}[h]
\caption{}
\label{alg1}
\begin{algorithmic}[1]
    \REQUIRE $\gamma\in (0,1)$ 
    \REQUIRE $x(k)\in \mathbb{R}$  state at initial time $k$.
    \REQUIRE $\varepsilon_0 \in (0,1)$
    \STATE $l \gets 0$
    \STATE $\varepsilon \gets \varepsilon_0$
    \STATE $(\hat{u}_c, p_c) \gets$ solve \textit{Problem~\ref{prb2}} 
    \WHILE{$\sum_{j=1}^{M_l}[1-g^{j}((10b),\hat{u}_c, p_c; \varepsilon)] > m_l$}
        \STATE $l \gets l + 1$
        \STATE $\varepsilon \gets \gamma \varepsilon$
        \STATE $(\hat{u}_c, p_c) \gets$ solve \textit{Problem~\ref{prb2}} 
    \ENDWHILE
    \RETURN ${\hat{u}(k)}^{*}, {p(k)}^{*}$
\end{algorithmic}
\end{algorithm}

From Lines 3 to 7 of \textbf{Algorithm~\ref{alg1}}, we see that our controller solves \textit{Problem~\ref{prb1}} approximately by alternating between solving \textit{Problem~\ref{prb2}} and verifying the chance-constraint. The approximation is improved through the repeated reduction of $\varepsilon$.

\begin{thm}\label{thr1}
Given confidence level $\delta \in (0, 1)$, suppose that \textbf{Algorithm~\ref{alg1}} terminates at $l^*$, then the solution \((\hat{u}(k)^{*},\ p(k)^{*})\) of the algorithm
 is feasible for \textit{Problem~\ref{prb1}} with probability at least \(1 - \delta\).
In other words, it holds that
\begin{equation}\label{verification4}
\Pr\left\{g({{(7b),(7c),(7d)},\hat{u}, p; {\gamma}^{l^{*}}{\varepsilon}_0)=1}\right\} > 1-\delta.
\end{equation}
\end{thm}
\vspace{3mm}
\textit{Theorem~\ref{thr1}} guarantees that the solution obtained by \textbf{Algorithm~\ref{alg1}} satisfies the chance-constraint (7b) with a confidence level of at least \(1 - \delta\) in addition to deterministic constraints (7c) and (7d).

\section{Numerical Experiments}\label{simu}
In this section, we present two experiments for demonstrating the design and operation of Gig-WMS. 
Experiment 1 is about constructing a realistic decision-making model by crowdsourcing. Experiment 2 is about demonstrating the verification algorithm.
\subsection{Experiment 1 Identification of the Utility Function via Crowdsourcing}  
In this subsection, we describe  Experiment 1, which focuses on the identification of the utility function (\ref{utility}) through crowdsourcing.  

\subsubsection{Data Collection}  
In Experiment 1, we conducted a survey on Yahoo! Crowdsourcing \cite{Yahoo} with 500 participants to investigate the relationship between the task hours and wages presented on the platform and the task acceptance probability. The survey questions are summarized below.   

\vspace{3mm}
\noindent\textbf{[Tokyo Residents Only]}\\
This is a survey on part-time gig-work during spare time. Please imagine food delivery tasks lasting up to 2 hours.
\begin{enumerate}
  \item If the working time is 30 minutes, what is the minimum wages (JPY) you would accept?\\
  \quad 500 \quad 550 \quad 600 \quad 650 \quad 700

  \item If the working time is 60 minutes, what is the minimum wages (JPY) you would accept?\\
  \quad 1000 \quad 1100 \quad 1200 \quad 1300 \quad 1400

  \item If the working time is 90 minutes, what is the minimum wages (JPY) you would accept?\\
  \quad 1500 \quad 1650 \quad 1800 \quad 1950 \quad 2100

  \item If the working time is 120 minutes, what is the minimum wages (JPY) you would accept?\\
  \quad 2000 \quad 2200 \quad 2400 \quad 2600 \quad 2800
\end{enumerate}
\vspace{3mm}
In this survey, we first provided an explanation to unify respondents' understanding of gig-work. Next, we included a disclaimer explaining the protection of personal information and confirming respondents' consent to participate. The survey consisted of four questions, each asking respondents to choose the minimum wages they would accept for a specific task hour from five options.  

By aggregating the survey results, we obtained the number of respondents (out of 500) who accepted the task under each combination of planned task hours $\hat{u}$ and wages $p$. Since the survey included four different task hours and five wage levels, we obtained data for 20 different conditions.  
Here, we define $z$ as the ratio of respondents who accepted the task among the 500 participants. Using these 20 data points $(\hat{u}, p, z)$, we estimated the coefficients $\kappa, \lambda, \nu$ of the utility function (\ref{utility}) by the least squares method. $\nu$ is the mean of $\nu_i$.

\subsubsection{Results}  
We now present the results of Experiment 1.  
The estimated values of $\kappa, \lambda, \nu$ obtained using the least squares method. As a result, $\kappa=-7.253,  \lambda=0.006385, \nu=-1.216$. We constructed the probability model in Fig. \ref{fit}. The blue surface represents the task acceptance probability, while the red points correspond to the aggregated survey results.  
\begin{figure}[t]
    \centering
    \includegraphics[scale=0.28]{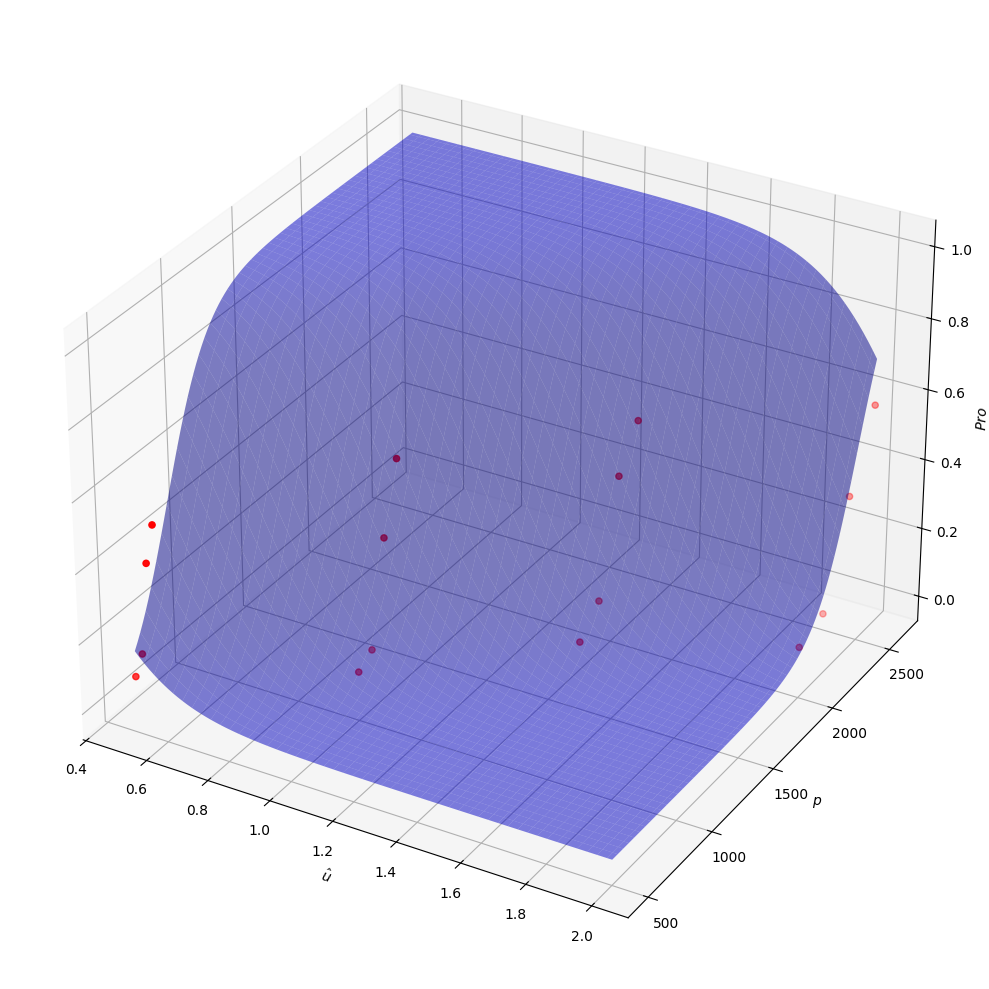}
    \caption{Surface of the task Acceptance Probability}
    \label{fit}
\end{figure}
From Fig. \ref{fit}, we observe that the task acceptance probability decreases as the planned task hours increase, and increases as the wages increase. Additionally, for a given planned task hour, the probability asymptotically approaches 1 as the wages increase.

\subsection{Experiment 2 Validation of the Feasibility Verification Algorithm}

In this subsection, we describe Experiment 2, which evaluates the effectiveness of the feasibility verification algorithm proposed in Section \ref{cont}. 

\subsubsection{Method}
In this experiment, we compare the control performance of two controllers to verify the effectiveness of the proposed feasibility verification algorithm, given in \textbf{Algorithm 1}. Among the two controllers, the one that uses the feasibility verification algorithm is referred to as Controller 1, and the one that does not is referred to as Controller 2.

Both controllers are based on the novel utility function (\ref{utility}) proposed in this paper. In other words, it is assumed that each gig-worker $i$ makes decisions according to a probability model (\ref{dec_utility}). The coefficients $\kappa$ and $\lambda$ are the values obtained in Experiment 1. $\nu_i$ is a uniformly distributed random variable whose mean is $\nu$. $\nu_i=[\nu-100\lambda, \nu + 100\lambda]$.

The numerical settings of the parameters are summarized in Table~\ref{table3}.
\begin{table}[t]
 \caption{Numerical Setting of Parameters Experiment 2}
 \label{table3}
 \centering
  \begin{tabular}{c p{5cm} c}
   \hline
   Parameter & Meaning & Setting Value \\
   \hline \hline
   $\kappa$ & Coefficient of $\hat{u}$ & -7.253 \\
   $\lambda$ & Coefficient of $p$  & 0.006385 \\
   $\nu$ & Constant term  & -1.216 \\
   $n$ & Number of gig-workers & 100 \\
   $x(0)$ & Initial value of $x$  & 30 \\
   $x_{\text{ref}}$ & Target value of $x$  & 10 \\
   $d$ & Additional workload & 5 \\
   $N$ & Prediction horizon & 3 \\
   $\alpha$ & Variable of the Riemann zeta function & 2 \\
   $\varepsilon_0$ & Initial value of $\varepsilon$ & 0.01 \\
   $\delta$ & Confidence level for feasibility verification & $10^{-8}$ \\
   $\eta$ & Tolerance probability for opportunity constraints & 0.05 \\
   \hline
  \end{tabular}
\end{table}
Based on the settings, we conducted 200 simulation runs and counted the number of cases in which the remaining workload $x(10)$ at last time $k = 10$ exceeded the target value $x_{\text{ref}}$, which implies that chance-constraint (7b) is violated, for both controllers. This was done to assess the effectiveness of the feasibility verification algorithm.

\subsubsection{Results}

Figures \ref{spv1} and \ref{spv2} show the histograms of the remaining workload $x(10)$ at last time $k = 10$ for Controllers 1 and 2, respectively. Fig. \ref{spv1} corresponds to the results from Controller 1, while Fig. \ref{spv2} shows the results from Controller 2.
\begin{figure}[t]
    \centering
    \begin{minipage}{0.48\linewidth}
        \centering
        \includegraphics[width=\linewidth]{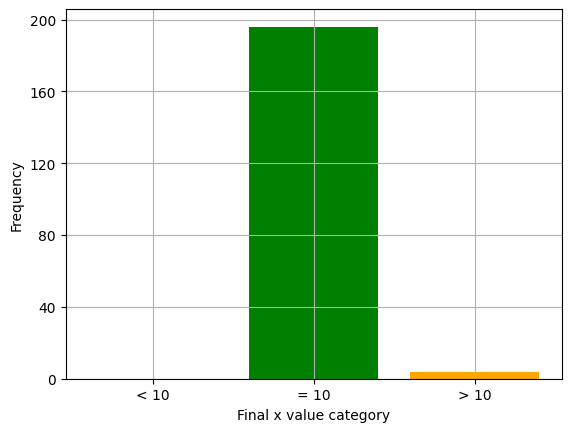}
        \caption{Histogram of $x(10)$ for Controller 1}
        \label{spv1}
    \end{minipage}
    \hfill
    \begin{minipage}{0.48\linewidth}
        \centering
        \includegraphics[width=\linewidth]{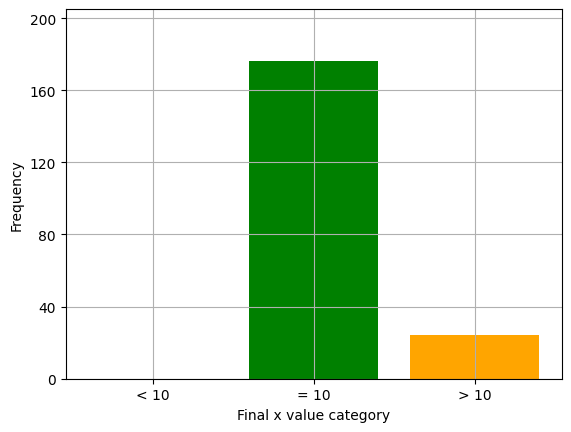}
        \caption{Histogram of $x(10)$ for Controller 2}
        \label{spv2}
    \end{minipage}
\end{figure}
As shown in Figs. \ref {spv1} and \ref{spv2}, the number of cases in which the remaining workload $x(10)$ exceeded the target value $x_{\text{ref}}$ was 4 for Controller 1, while it was 24 for Controller 2. Given that $\eta = 0.05$, the allowable number of failures among 200 simulations is approximately 10.

These results show that incorporating the feasibility verification algorithm into the controller enables the generation of task offers that satisfy feasibility with the desired confidence level.

\section{Conclusion and Future Work}\label{conclu}
In this paper, we constructed a decision-making model of a group of gig-workers who make probabilistic decisions, based on real data collected from surveys.  
As an application of this model, we designed an efficient work management system for a group of gig-workers (Gig-WMS).  
Additionally, we conducted simulations to verify whether Gig-WMS performs as expected. The results showed that the system meets the desired control performance set by the system designer.

\bibliography{ifacconf}  

\end{document}